\documentclass[twocolumn,pra,showpacs]{revtex4}

\usepackage{amsmath}
\usepackage{graphicx}
\usepackage{amsfonts}
\usepackage{amssymb}
\newcommand{\beq}{\begin{equation}}
\newcommand{\eeq}{\end{equation}}
\newcommand{\beqr}{\begin{eqnarray}}
\newcommand{\eeqr}{\end{eqnarray}}

\newcommand{\ct}[1]{\cite{#1}}
\newcommand{\dpr}{\delta P_r}
\newcommand{\dpd}{\delta P_2}
\newcommand{\da}{\delta A}
\newcommand{\dain}{\delta A^{in}}

\begin{document}
\title{Manipulation and storage of optical field and atomic ensemble quantum states}

\author{A. Dantan\footnote[3]{email: dantan@spectro.jussieu.fr}}

\author{A. Bramati}

\author{M. Pinard}

\author{E. Giacobino}

\affiliation{Laboratoire Kastler Brossel, Universit\'{e} Pierre et
Marie Curie,\\
Case 74, 4 place Jussieu, 75252 Paris Cedex 05, France}

\begin{abstract}
We study how to efficiently manipulate and store quantum
information between optical fields and atomic ensembles. We show
how various non-dissipative transfer schemes can be used to
transfer and store quantum states such as squeezed vacuum states
or entangled states into the long-lived ground state spins of
atomic ensembles.
\end{abstract}


\maketitle

\section{Introduction}

If photons are known to be fast and robust carriers of quantum
information, a major difficulty is to store their quantum state.
In the continuous variable regime a number of non-classical
optical field states - squeezed or entangled states - have been
generated with great efficiency
\cite{grangier,lambrecht,bowen1,silberhorn,josse1,bowen,glockl,josse,laurat}.
However, in order to realize scalable quantum networks
\cite{bennett} quantum memory elements are required to store and
retrieve optical field states. To this end atomic ensembles have
been widely studied as potential quantum memories \cite{lukin}.
Indeed, the long-lived collective spin of an atomic ensemble with
two ground state sublevels appears as a good candidate for the
storage and manipulation of quantum information conveyed by light
\cite{zoller}. Various schemes have already been studied: first,
the recent "slow-" and "stopped-light" experiments have shown that
it was possible to store a light pulse inside an atomic cloud
\cite{hau,phillips} in the Electromagnetically Induced
Transparency (EIT) configuration \cite{harris}. EIT is known to
occur when two fields are both one- and two-photon resonant with
3-level $\Lambda$-type atoms, which allows one field to propagate
without dissipation through the medium. However, the storage has
only been demonstrated for
classical variables so far.\\
On the other hand, the stationary mapping of a quantum state of
light (squeezed vacuum) onto an atomic ensemble, as well as the
conditional entanglement of two ensembles, have been
experimentally demonstrated, this time in an off-resonant Raman
configuration \cite{julsgaard} and in a single pass scheme.
Quantum state transfers between light and atoms are also
interesting in relation to "spin squeezing" and high precision
measurements
\cite{wineland} and have been widely studied \cite{bigelow,kozhekin,vernac,molmer,dantan1}.\\
In the present paper we present a model for the interaction
between optical fields in cavity and atomic ensembles, and show
various examples of non-destructive atom-field quantum state
transfers \cite{dantan3,dantan4}. In the first Section we show how
to write an optical field quantum state - a squeezed vacuum state
- onto the ground state coherence of an atomic ensemble. We assess
the efficiency of the mapping in different situations, as well as
the storage into the atoms. We then consider the reverse transfer
operation, from the atoms to the field, and show that it is
possible to perform a quasiperfect readout of the atomic state in
the field exiting the cavity. In the next Section we show how
these results extend to the manipulation and storage of
Einstein-Podolsky-Rosen (EPR) entangled states. In the last
Section we study how to transfer the squeezing stored into one
ensemble into a second.

\section{Quantum state transfer between field and atoms}

\subsection{Model system and evolution equations}

\begin{figure}[h]
  \centering
  \includegraphics[width=5cm]{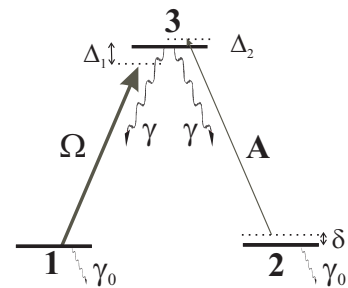}
  \caption{Three-level system in a $\Lambda$ configuration.}\label{fig1}
\end{figure}

The interaction considered throughout this paper is schematically
represented in Fig. \ref{fig1}: a set of $N$ $3$-level atoms in a
$\Lambda$ configuration interacts on each transition with one mode
of the electromagnetic field in an optical cavity. The $3$-level
system can be described using $9$ collective operators for the $N$
atoms of the ensemble: the populations $\Pi _{i}=\sum\limits_{\mu
=1}^{N}\left| i\right\rangle _{\mu }\left\langle i\right| _{\mu }$
($i=1-3)$, the components of the optical dipoles $P_{i}$ in the
frames rotating at the frequency of their corresponding lasers and
their hermitian conjugates and the components of the dipole
associated to the ground state coherence: $ P_{r}=\sum\limits_{\mu
=1}^{N}\left| 2\right\rangle _{\mu }\left\langle 1\right| _{\mu }$
and $P_{r}^{\dagger }$.
The atom-field coupling constants are defined by $g_i={\cal
E}_{0i}d_i/\hbar $, where $d_i$ are the atomic dipoles, and ${\cal
E}_{0i}=\sqrt{ \hbar \omega _{i}/2\epsilon _{0}{\cal S}c}$ (${\cal
S}$ being the beam cross-section). With this definition, the mean
square value of a field is expressed in number of photons per
second. To simplify, the decay constants of dipoles $P_{1}$ and
$P_{2}$ are both equal to $\gamma$. In order to take into account
the finite lifetime of the two ground state sublevels $1$ and $2$,
we include in the model another decay rate $\gamma _{0}$, which is
supposed to be much smaller than $\gamma $. We also consider that
the sublevels $1$ and $2$ are repopulated with in-terms $\Lambda
_{1}$ and $\Lambda _{2}$, so that the total atomic population is
kept constantly equal to $N$.

The evolution of such a system is given by a set of quantum
Heisenberg-Langevin equations

\begin{eqnarray}
\dot{\Pi}_{1} &=&i\Omega^* P_{1}-i\Omega P_{1}^{\dagger }+\gamma
\Pi _{3}-\gamma
_{0}\Pi _{1}+\Lambda _{1}+F_{11}  \label{pi1} \nonumber\\
\dot{\Pi}_{2} &=&ig_2A^{\dagger }P_{2}-ig_2AP_{2}^{\dagger
}+\gamma \Pi _{3}-\gamma _{0}\Pi _{2}+\Lambda
_{2}+F_{22}  \label{pi2} \nonumber\\
\dot{\Pi}_{3} &=&-(i\Omega^*P_{1}-i\Omega P_{1}^{\dagger
})-(ig_2A^{\dagger }P_{2}-ig_2AP_{2}^{\dagger
})\nonumber\\
&&\hspace{3cm}-2\gamma\Pi _{3}+F_{33}  \label{pi3} \nonumber\\
\dot{P}_{1} &=&-(\gamma +i\Delta _{1})P_{1}+i\Omega(\Pi
_{1}-\Pi _{3})+ig_2AP_{r}^{\dagger }+F_{1}  \label{p13} \nonumber\\
\dot{P}_{2} &=&-(\gamma +i\Delta _{2})P_{2}+ig_2A(\Pi _{2}-\Pi
_{3})+i\Omega P_{r}+F_{2}  \label{p23} \nonumber\\
\dot{P}_r&=&-(\gamma_0-i\delta)P_r+i\Omega^*P_2-ig_2AP_1^{\dagger}+f_{r}\label{pr}\nonumber\\
\dot{\Omega} &=&-(\kappa +i\Delta _{c1})\Omega+\frac{ig_1^2}{\tau
} P_{1}+\sqrt{\frac{2\kappa }{\tau
}}\Omega^{in}  \label{A1}\nonumber\\
\dot{A} &=&-(\kappa +i\Delta _{c2})A+\frac{ig_2}{\tau }
P_{2}+\sqrt{\frac{2\kappa }{\tau }}A^{in}  \nonumber\label{A2}
\end{eqnarray}

where the $g_i$'s are assumed real, $\Omega$ is the Rabi frequency
associated to the control field, $\delta =\Delta _{1}-\Delta _{2}$
is the two-photon detuning, $\kappa $ is the intracavity field
decay and $\tau $ the round trip time in the cavity, so that
$T=2\kappa\tau$ represents the transmission of the cavity coupling
mirror. The $F$'s are standard $\delta$-correlated Langevin
operators taking into account the coupling with the other cavity
modes. From the previous set of equations, it is possible to
derive the steady state values and the correlation matrix for the
fluctuations of the whole atom-field system (see e.g.
\cite{vernac}). In the following Sections we look for the best
regimes for efficient quantum state transfers between fields and
atoms and derive simplified equations for the transfer processes.

\subsection{Decoupled equations for the fluctuations}

We consider a very simple situation in which field $\Omega$ plays
the role of a control parameter and field $A$ has zero mean value.
In this case all the atoms are pumped in $|2\rangle$, so that only
$\langle\Pi_2\rangle$ is non zero in steady state. The
fluctuations for $\delta P_r$, $\delta P_2$ and $\delta A$ are
then decoupled from the other operators fluctuations
\begin{eqnarray} \dot{\dpr}&=&-(\gamma_0-i\delta)\dpr+i\Omega \dpd+f_r\label{dpr}\\
\dot{\dpd}&=&-(\gamma+i\Delta)\dpd+i\Omega \dpr+igN\da+F_{2}\label{dpd}\\
\delta\dot{A}&=&-(\kappa+i\Delta_c)\da+\frac{ig}{\tau}\dpd+\sqrt{\frac{2\kappa}{\tau}}\dain\label{da},\end{eqnarray}
which allows analytical calculations and simple physical
interpretations. The atomic spin associated to the ground states
is aligned along $z$ at steady state: $\langle J_z\rangle=\langle
\Pi_2-\Pi_1\rangle/2=N/2$. The spin quantum state is then given by
the coherence components, $J_x=(P_r+P_r^{\dagger})/2$ and
$J_y=(P_r-P_r^{\dagger})/2i$. Their commutator,
$[J_x,J_y]=iJ_z=iN/2$, is then very similar to that of the field
quadrature operators: $[X_{\varepsilon},Y_{\varepsilon}]=2i$,
where
$X_{\varepsilon}=Ae^{-i\varepsilon}+A^{\dagger}e^{i\varepsilon}$
and $Y_{\varepsilon}=X_{\varepsilon+\pi/2}$. The field or the
atomic quantum state can be represented in a symmetrical fashion
by the noise ellipsoid in the conjugate variable plane. For
instance, as the field is said to be squeezed when the noise of
one quadrature $X_{\varepsilon}$ is less than the shot-noise value
of 1, the spin component $J_{\theta}=J_x\cos\theta +J_y\sin\theta$
in the ($x,y$)-plane is said to be spin-squeezed when its variance
is less than the coherent state value $|\langle J_z\rangle|/2$,
and the degree of spin-squeezing is given by \cite{ueda}
\begin{eqnarray} \Delta J^2_{min}=\min_{\theta}\frac{\Delta
J^2_{\theta}}{|\langle J_z\rangle|/2}<1.\end{eqnarray} We now
explicit schemes in which quantum states (squeezed or entangled
states) can be transferred between field and atoms in this
representation.

\subsection{Writing onto atoms}\label{writing}

For such a system two situations are particularly interesting for
non-dissipative transfer processes: one is the so-called
Electromagnetically Induced Transparency configuration, in which
the fields are both one- and two-photon resonant
($\Delta=\delta=0$). The other is the Raman configuration, where
the one-photon detuning is much larger than the exited state
linewidth ($\Delta\gg\gamma$). These two interactions are rather
insensitive to spontaneous emission and, therefore, very favorable
to non-destructive quantum state transfer operations. For
relatively bad cavities the field and the optical coherences
evolve rapidly as compared to the ground state coherence. In this
limit it is possible to adiabatically eliminate these operators
and derive simple analytical equations for the ground state
observables:
\begin{eqnarray} \label{spin1}
\delta\dot{J}_x &=&-\gamma_{\varepsilon}\delta
J_{x}-\beta_{\varepsilon} \delta X_{\varepsilon}^{in}+\tilde{f}_{x}\\
\delta\dot{J}_y&=&-\gamma_{\varepsilon}\delta
J_{y}-\beta_{\varepsilon} \delta
Y_{\varepsilon}^{in}+\tilde{f}_{y}\label{spin2}
\end{eqnarray}
where $\varepsilon$ stands for the situation considered: "EIT"
($\Delta=0$) is denoted by $\varepsilon=0$, whereas the "Raman"
configuration ($\Delta\gg\gamma$) corresponds to
$\varepsilon=\pi/2$. With these notations
$\gamma_{\varepsilon}=\gamma_0+\Gamma_{\varepsilon}$ is the
effective pumping rate for the fluctuations
[$\Gamma_{0}=\Gamma_E/(1+2C)$ in EIT and
$\Gamma_{\pi/2}=(1+2C)\Gamma_R$ in a Raman configuration].
$\beta_{\varepsilon}$ is the coupling coefficient between the
incident field and the atomic coherence
[$\beta_0=gN\Omega/\gamma(1+2C)\sqrt{T}$ and
$\beta_{\pi/2}=gN\Omega/\Delta \sqrt{T}$], and the $\tilde{f}$'s
are effective quantum Langevin operators accounting for
dissipation. In both cases the effective two-photon detuning and
the effective cavity detuning are set to 0. Both situations derive
from formally similar effective Hamiltonians
\begin{eqnarray} H_{\varepsilon}=\hbar\frac{2\beta_{\varepsilon}}{N}\left[J_xY_{\varepsilon}^{in}-
J_yX_{\varepsilon}^{in}\right].\end{eqnarray} Since this
Hamiltonian is nothing but the coupling between two harmonic
oscillators, the physical interpretation is clear: if one knows
the input field state $A^{in}$ variance matrix one can deduce that
of the atomic spin. In EIT, for instance, and for a broadband
squeezed vacuum input, the spin-squeezed component angle
$\theta_{sq}$ will be that of the field squeezed quadrature:
$\theta_{sq}=\varepsilon_{sq}$. In a Raman configuration a $\pi/2$
rotation of the spin should be performed:
$\theta_{sq}=\varepsilon_{sq}+\pi/2$. In both cases the minimum
variance takes a similar form
\begin{equation} \label{var}\Delta
J^2_{min}=\frac{2C}{1+2C}\frac{\Gamma_{\varepsilon}}{\gamma_{\varepsilon}}e^{-2r}+
\frac{\gamma_0}{\gamma_{\varepsilon}}+\frac{1}{1+2C}\frac{\Gamma_{\varepsilon}}{\gamma_{\varepsilon}}\end{equation}
where $e^{-2r}$ is the incident field squeezing. The first term
($\varpropto e^{-2r}$) reflects the incident field state, the
second ($\varpropto \gamma_0$) is the noise contribution
associated to the loss of ground state coherence and the third
($\varpropto \Gamma_{\varepsilon}$) is the noise contribution
coming from spontaneous emission.
\begin{figure}[h]
  \centering
  \includegraphics[width=7cm]{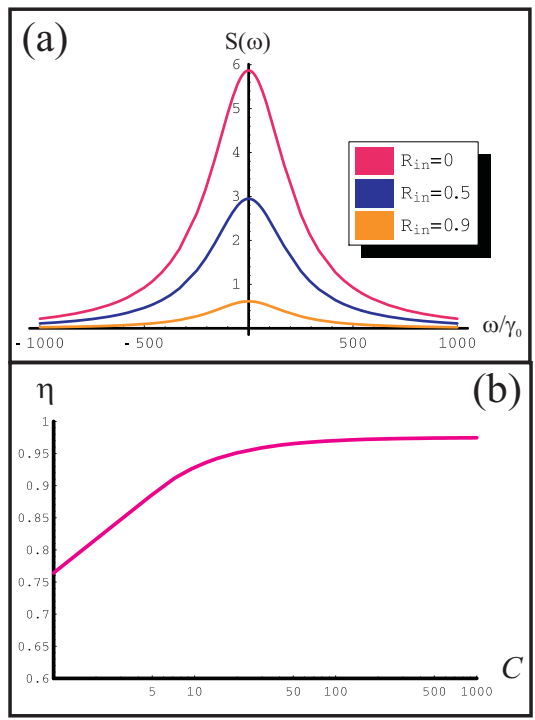}
  \caption{(a) Atomic noise spectra for the least noisy spin component,
  for different values of the input squeezing $R_{in}=1-e^{-2r}$.
  The effective pumping rate and the cooperativity are the same
  for each curve ($C=100,\gamma_{\varepsilon}=0.075\gamma$).
  (b) Optimized transfer efficiency versus cooperativity.}\label{fig2}
\end{figure}
Consequently, one sees that a good quantum state transfer -
$\Delta J^2_{min}\sim e^{-2r}$ - occurs in the regime $C\gg 1$ and
$\gamma_0\ll \Gamma_{\varepsilon}\ll \gamma,\kappa$. A useful
quantity to characterize the quality of the quantum state transfer
is provided by the transfer efficiency $\eta=(1-\Delta
J^2_{min})/(1-e^{-2r})$, which can be written as
\begin{equation}
\eta_{\varepsilon}=\frac{2C}{1+2C}\frac{\Gamma_{\varepsilon}}{\gamma_0+\Gamma_{\varepsilon}}\end{equation}
In Fig. \ref{fig2}(b) we show the transfer efficiency versus the
cooperativity. For each value of $C$ the optical pumping was
optimized numerically in order to maximize the efficiency
\ct{dantan3}. A high efficiency is possible for rather small
values of the cooperativity.

Note that, in this case, one could also define a standard quantum
limit for the atomic noise by looking at the atomic noise
spectrum. In this low-frequency approximation the atomic coherence
noise spectra have a Lorentzian shape with FHWM given by
$2\gamma_{\varepsilon}$. For the squeezed component this
Lorentzian has a peak value decreased by a factor $e^{-2r}$ as
compared to that of the corresponding coherent state [see Fig.
\ref{fig2}(a)]. However, this notion of standard quantum limit at
a given frequency is only relevant when comparing with the noise
spectrum of a coherent spin state under the same conditions (same
pumping strength $\gamma_{\varepsilon}$, same number of atoms...).
Besides, $\gamma_{\varepsilon}$ represents the quantum memory
storage frequency bandwidth. An interesting feature of this cavity
scheme is that it is much broader than the natural linewidth
$\gamma_0$.

This simplified model can be shown to be in excellent agreement
with full quantum calculations in the regime considered. A more
detailed study of what happens when $\Gamma_{\varepsilon}$ is
increased, when the detunings are non-zero or when arbitrary field
states for $A$ are used can be found in Ref. \ct{dantan3}.

\subsection{Storage and readout}\label{readout}

The squeezing transfer can be considered completed after a time of
a few $1/\gamma_{\varepsilon}$. If all fields are abruptly
switched off one is left with a spin squeezed atomic ensemble. The
atomic squeezing decays very slowly on a timescale given by
$1/\gamma_0$. After a storage time $t_s$, small with respect to
this decay time, one can retrieve the atomic state into the field
exiting the cavity by switching on again \textit{only} the control
field. Indeed, neglecting $\gamma_0$ and in the regime
$\Gamma_{\varepsilon}\ll\gamma,\kappa$ the outgoing field mode can
be shown to be \beqr\label{Xout} \delta
X_{\varepsilon}^{out}(t)&=& \delta X_{\varepsilon}^{in}(t)-\alpha
\delta J_{\varepsilon}(0)e^{-\Gamma_{\varepsilon}t}\\\nonumber
&&-2\eta^2[ \delta
X_{\varepsilon}^{in}(t)-\Gamma_{\varepsilon}\int_0^t
e^{-\Gamma_{\varepsilon}(t-s)}\delta X_{\varepsilon}^{in}(s)ds]
\\\nonumber &&+\beta[ \delta X_{v\varepsilon}(t)-\Gamma_{\varepsilon}\int_0^t
e^{-\Gamma_{\varepsilon}(t-s)} \delta X_{v\varepsilon}(s)ds],\eeqr
with $\alpha=\eta\sqrt{8\Gamma_{\varepsilon}/N}$,
$\beta=2\eta/\sqrt{1+2C}$ and $X_{v\varepsilon}$ is a white noise
operator corresponding to a normalized Langevin operator with
unity spectrum. $\eta=2C/(1+2C)$ is the efficiency for
$\gamma_0=0$, independent of the interaction considered. The terms
in $X_{\varepsilon}^{in}$, $X_{v\varepsilon}$ are intrinsic and
added field noise terms, whereas the term in $J_{\varepsilon}$
provides the quantum information relative to the incident atomic
state. The two-time correlation function has a much simpler form
\begin{eqnarray} \label{correlation}\mathcal{C}(t,t')&\equiv &\langle\delta
X_{\varepsilon}^{out}(t)\delta
X_{\varepsilon}^{out}(t')\rangle\\&=&\delta(t-t')-\frac{2C}{1+2C}2\Gamma_{\varepsilon}\left[1-\Delta^2
J_{\varepsilon}(0)\right]e^{-\Gamma_{\varepsilon}(t+t')}\nonumber\end{eqnarray}
In the absence of coupling [$\Gamma_{\varepsilon}=0$], or for a
coherent spin state [$\Delta J_{\varepsilon}^2=1$], one naturally
retrieves a shot-noise limited free field, with a
$\delta$-correlation function. However, if the atoms are
spin-squeezed one transitorily observes sub-shot noise
fluctuations for the outgoing field. It was shown in
\cite{dantan3} that it is possible to measure the atomic state
with almost $100\%$ efficiency using a homodyne detection and
correctly choosing the local oscillator temporal profile - or,
equivalently, by choosing the right electronic gain in the
detection process. Using an optimal matching in
$e^{-\Gamma_{\varepsilon}t}$, adapted to the atomic temporal
response, the readout efficiency, defined as the ratio of field
squeezing to atomic squeezing (at switching time), is then also
given by $\eta$. Taking into account that the atomic squeezing has
decreased by a factor $e^{-2\gamma_0t_s}$ during storage, the
global efficiency of the quantum memory is then
$\eta^2e^{-2\gamma_0t_s}$.

\section{EPR-correlated atomic ensembles}\label{ensembles}

\begin{figure}[h]
  \centering
  \includegraphics[width=6cm]{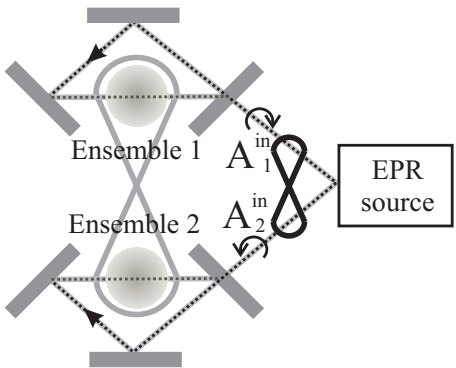}
  \caption{Scheme for entanglement storage into two ensembles.}\label{fig3}
\end{figure}
In this Section we show how to generalize the previous quantum
state transfer to quantum correlated states, or EPR states, which
are of great importance in many quantum information protocols in
the continuous variable regime. As mentioned earlier such states
are now readily produced by different sources and with very good
efficiency \ct{bowen,glockl,josse,laurat}. We therefore assume
that we dispose of a pair of EPR-entangled vacuum fields, $A_1$
and $A_2$, and of a pair of identical ensembles (1) and (2), as
shown in Fig. \ref{fig3}. The amount of EPR-type correlations
between the incident field modes is quantified using the
inseparability criterion \cite{duan}
\begin{equation}\label{duanfield}
\mathcal{I}_{f}^{in}=\frac{1}{2}[\Delta^2(X_1^{in}-X_2^{in})+\Delta^2(Y_1^{in}+Y_2^{in})]<2\end{equation}
For the spins $J_{x1}-J_{x2}$ and $J_{y1}+J_{y2}$ are the
equivalent of the EPR operators, since $\langle
[J_{x1}-J_{x2},J_{y1}+J_{y2}]\rangle=i\langle
J_{z1}-J_{z2}\rangle=0$ when spins 1 and 2 are equal and parallel.
A similar criterion to (\ref{duanfield}) can be derived for the
inseparability of spins 1 and 2
\begin{equation}\nonumber
\Delta^2(J_{x1}-J_{x2})+\Delta^2(J_{y1}+J_{y2})<|\langle
J_{z1}\rangle|+|\langle J_{z2}\rangle|=N\end{equation} As in the
previous Section "EIT"- or "Raman"-type interactions with both
ensembles lead to coupling between the incident EPR-fields and the
spin coherence components
\begin{eqnarray} \frac{d}{dt}(\delta J_{x1}-\delta
J_{x2})&=&-\gamma_{\varepsilon}(\delta J_{x1}-\delta
J_{x2})\\&&-\beta_{\varepsilon} (\delta
X_{1\varepsilon}^{in}-\delta X_{2\varepsilon}^{in})+
\tilde{f}_{x1}-\tilde{f}_{x2}\nonumber\\
\frac{d}{dt}(\delta J_{y1}+\delta
J_{y2})&=&-\gamma_{\varepsilon}(\delta J_{y1}+\delta
J_{y2})\\&&-\beta_{\varepsilon} (\delta
Y_{1\varepsilon}^{in}+\delta
Y_{2\varepsilon}^{in})+\tilde{f}_{y1}+\tilde{f}_{y2}\nonumber
\end{eqnarray}
and one can show that the field entanglement is efficiently
transferred to the spins \cite{dantan4}
\begin{equation}\label{simple}\mathcal{I}_{at}=\frac{2C}{1+2C}\frac{\Gamma_{\varepsilon}}{\gamma_{\varepsilon}}
\;\mathcal{I}_{f}+2\left[\frac{\gamma_0}{\gamma_{\varepsilon}}+\frac{\Gamma_{\varepsilon}}{(1+2C)\gamma_{\varepsilon}}\right]\end{equation}
where
$\mathcal{I}_{at}=\frac{2}{N}[\Delta^2(J_{x1}-J_{x2})+\Delta^2(J_{y1}+J_{y2})]$
stands for the atomic entanglement (normalized to 2). The same
conclusions hold: for a good cooperative behavior ($C\gg 1$) and
for $\gamma_0\ll \gamma_{\varepsilon}\ll \gamma,\kappa$, the added
noise terms ($\varpropto 1/(1+2C)$ and $\gamma_0$) are negligible
compared to the coupling, and the atomic entanglement
is close to the initial field entanglement, $\mathcal{I}_{at}\sim\mathcal{I}_f$.\\
The same readout scheme as previously can be applied to retrieve
entanglement in a transient manner between the outgoing fields.
This entanglement can be measured using the techniques developed
in Refs. \cite{josse,dantan3,josse2}. Note that the lifetime of
this entanglement is given by the phenomenological time constant
$1/\gamma_0$ introduced in our model. For cold atoms it can
represent the loss of atoms out of the trap, and for atomic vapors
the depolarizing time. We have neglected the collisions leading to
a depolarization of the spin, which is legitimate for cold atoms,
but should be considered for vapors if one were to evaluate
precisely the storage time of the quantum memory.

\section{Pseudo-quantum repeater}

\begin{figure}[h]
  \centering
  \includegraphics[width=8cm]{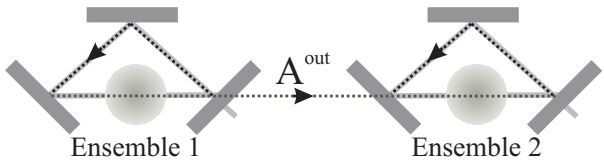}
  \caption{Scheme for the "pseudo" quantum repeater.}\label{fig4}
\end{figure}
We assume that we dispose of two identical atomic ensembles [Fig.
\ref{fig4}] and that, using the techniques of Sec. \ref{writing},
we have spin-squeezed ensemble 1 to some degree $e^{-2r_1}$ on the
$x$-component and that it is in a minimum uncertainty state
($\Delta J_{y1}^2=e^{2r_1}$). Spin 2 is initially in a coherent
spin state aligned along $z$. If we perform an optical readout of
ensemble 1 by switching on the control field in the first cavity
the outgoing field is squeezed, as can be seen from
(\ref{correlation}). It can then be used as input for the spin in
the second cavity \beqr\label{dotjx2} \delta
\dot{J}_{x2}&=&-\gamma_{\varepsilon}\delta
J_{x2}-\beta_{\varepsilon}\delta
X_{\varepsilon}^{out}+\tilde{f}_{x2}\\\label{dotjy2} \delta
\dot{J}_{y2}&=&-\gamma_{\varepsilon}\delta
J_{y2}-\beta_{\varepsilon}\delta
Y_{\varepsilon}^{out}+\tilde{f}_{y2}\eeqr  where
$X^{out}_{\varepsilon}$ and $Y^{out}_{\varepsilon}$ are input
fields out of the first cavity, the expression of which is given
by (\ref{Xout}). In the previous equations we have neglected the
transit time from one cavity to the other. The variances of spin 2
coherence components can be calculated from Eqs. (\ref{Xout}),
(\ref{dotjx2}) and (\ref{dotjy2}); one gets, after normalization
by the atomic shot-noise $N/4$,
\begin{eqnarray} \Delta J_{x2}^2(t)&=&
1-\eta^4(2\gamma_{\varepsilon}t)^2e^{-2\gamma_{\varepsilon}t}[1-e^{-2r_1}]\\
\Delta J_{y2}^2(t)&=&
1+\eta^4(2\gamma_{\varepsilon}t)^2e^{-2\gamma_{\varepsilon}t}[e^{2r_1}-1]\eeqr
The squeezing in the second cavity is maximum for
$t=1/\gamma_{\varepsilon}$ and is related to the squeezing in the
first cavity: \beqr 1-e^{-2r_2}=\frac{4}{e^2}\eta^4
[1-e^{-2r_1}]\eeqr Since $4/e^2\simeq 0.54$, only a little bit
more than half of the initial squeezing can be transferred to the
second ensemble with direct method.
\begin{figure}[h]
  \centering
  \includegraphics[width=8cm]{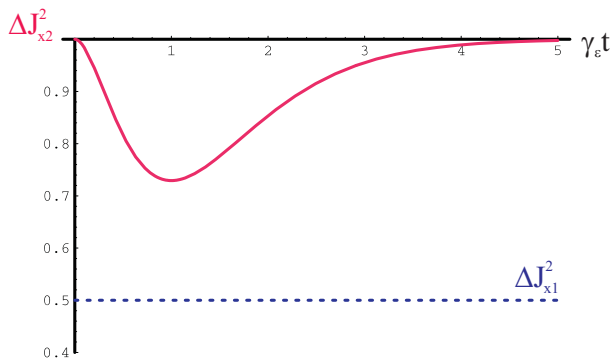}
  \caption{Minimal variance of spin 2 as a function of time: the squeezing increases to a
  maximum for $\gamma_{\varepsilon}t\simeq 1$ and then decreases back to 0.
  The initial squeezing in ensemble 1 is $1-e^{-2r_1}=0.5$.}\label{fig5}
\end{figure}
Another way to understand this imperfect transfer is to go to the
Fourier domain and see that the first ensemble response has a
spectral width given by $\gamma_{\varepsilon}$, and the input
squeezing spectrum of $X^{out}$ is itself multiplied by the same
response function for spin 2. The atomic noise spectrum in the
second cavity is then the product of two Lorentzian profiles with
equal width $\gamma_{\varepsilon}$. The atomic noise is then at
best the integral of this squared Lorentzian, which results in
this approximately 50\% quantum state transfer. Note that having
different widths for the readout of spin 1 and the writing on spin
2 does not improve the result. To fully transfer the state of spin
1 to spin 2 one actually needs a more refined protocol,
\textit{atomic teleportation} \cite{kuzmich,dantan5}, which
requires entanglement of the kind used in Sec. \ref{ensembles}.

\section{Conclusion}

We have presented a quantum model in which non-dissipative
interactions provide quasiperfect quantum state transfer between
optical fields and atomic ensembles spins. Field squeezed states
and EPR-entangled states can be stored with high efficiency into
atoms for a long time, and read out at will in the fields exiting
the cavities. Since both the squeezing and the entanglement are
conserved in such operations these results should be of importance
for the realization of robust quantum information and
communication networks involving optical fields and atomic
ensembles. Last, we examined the possibility to transfer by these
techniques the squeezing of a first atomic ensemble to a second.
However, the efficiency of the second mapping is limited to about
50\%. To achieve a perfect mapping one needs to perform a full
quantum teleportation protocol. This can be done by combining all
the ingredients presented in this paper \ct{dantan5}.

\end{document}